 \documentclass[final,5p,times,twocolumn]{elsarticle}

 \usepackage{graphics}
 \usepackage{graphicx}
\usepackage{amssymb}
\usepackage{amsmath}

\usepackage{xcolor}

\usepackage{textcomp}

\journal{Journal of Molecular Spectroscopy}

\begin{document}

\begin{frontmatter}

%% Title, authors and addresses

%% use the tnoteref command within \title for footnotes;
%% use the tnotetext command for the associated footnote;
%% use the fnref command within \author or \address for footnotes;
%% use the fntext command for the associated footnote;
%% use the corref command within \author for corresponding author footnotes;
%% use the cortext command for the associated footnote;
%% use the ead command for the email address,
%% and the form \ead[url] for the home page:
%%
%% \title{Title\tnoteref{label1}}
%% \tnotetext[label1]{}
%% \author{Name\corref{cor1}\fnref{label2}}
%% \ead{email address}
%% \ead[url]{home page}
%% \fntext[label2]{}
%% \cortext[cor1]{}
%% \address{Address\fnref{label3}}
%% \fntext[label3]{}

\title{Millimeter and submillimeter wave spectroscopy of propanal}

%% use optional labels to link authors explicitly to addresses:
%% \author[label1,label2]{<author name>}
%% \address[label1]{<address>}
%% \address[label2]{<address>}

\author[Koeln]{Oliver Zingsheim\corref{cor1}}
\ead{zingsheim@ph1.uni-koeln.de}
\cortext[cor1]{Corresponding author.}
\author[Koeln]{Holger S.P.~M\"uller}
\author[Koeln]{Frank Lewen}
\author[Kopenhagen]{Jes K. J{\o}rgensen}
\author[Koeln]{Stephan Schlemmer}

\address[Koeln]{I.~Physikalisches Institut, Universit{\"a}t zu K{\"o}ln, 
                Z{\"u}lpicher Str. 77, 50937 K{\"o}ln, Germany}
\address[Kopenhagen]{Centre for Star and Planet Formation, Niels Bohr Institute and Natural History Museum of Denmark, 
                     University of Copenhagen, {\O}ster Voldgade 5$-$7, 1350 Copenhagen K, Denmark}

%%%%%%%%%%%%%%%%%%%%%%%%%%%%%%%%%%%%%%%%%%%%%%%%%%%%%%%%%%%%%%%%%%%%%%%%%%%%%%%%%%%%%
%%%%%%%%%%%%%%%%%%%%%%%%%%%%%%%%%%%%%%%%%%%%%%%%%%%%%%%%%%%%%%%%%%%%%%%%%%%%%%%%%%%%%
%%%%%%%%%%%%%%%%%%%%%%%%%%%%%%%%%%%%%%%%%%%%%%%%%%%%%%%%%%%%%%%%%%%%%%%%%%%%%%%%%%%%%
\begin{abstract}

The rotational spectra of the two stable conformers \textit{syn}- and \textit{gauche}-propanal (CH$_3$CH$_2$CHO) were studied in the 
millimeter and submillimeter wave regions from 75 to 500$\,$GHz with the Cologne (Sub-)Millimeter wave Spectrometer. 
Furthermore, the first excited states associated with the aldehyde torsion and with the methyl torsion, respectively, of the \textit{syn}-conformer were analyzed. 
The newly obtained spectroscopic parameters yield better predictions, thus fulfill sensitivity and resolution requirements 
in new astronomical observations in order to unambiguously assign pure rotational transitions of propanal. This is demonstrated on a radio astronomical spectrum from the Atacama Large Millimeter/submillimeter Array Protostellar Interferometric Line Survey (ALMA-PILS). 
In particular, an accurate description of observed splittings, caused by internal rotation of the methyl group 
in the \textit{syn}-conformer and by tunneling rotation interaction from two stable degenerate \textit{gauche}-conformers, 
is reported. The rotational spectrum of propanal is of additional interest because of its two large amplitude motions 
pertaining to the methyl and the aldehyde group, respectively.

\end{abstract}

\begin{keyword}  %%% up to 6 !!
%% keywords here, in the form: keyword \sep keyword

propanal \sep 
rotational spectroscopy \sep 
submillimeter wave spectroscopy \sep 
interstellar molecule \sep 
internal rotation \sep
tunneling rotation interaction

%% MSC codes here, in the form: \MSC code \sep code
%% or \MSC[2008] code \sep code (2000 is the default)

\end{keyword}

\end{frontmatter}

%%
%% Start line numbering here if you want
%%
% \linenumbers

%%%%%%%%%%%%%%%%%%%%%%%%%%%%%%%%%%%%%%%%%%%%%%%%%%%%%%%%%%%%%%%%%%%%%%%%%%%%%%%%%%%%%
%%%%%  main text  %%%%%%%%%%%%%%%%%%%%%%%%%%%%%%%%%%%%%%%%%%%%%%%%%%%%%%%%%%%%%%%%%%%
%%%%%%%%%%%%%%%%%%%%%%%%%%%%%%%%%%%%%%%%%%%%%%%%%%%%%%%%%%%%%%%%%%%%%%%%%%%%%%%%%%%%%

%%%%%%%%%%%%%%%%%%%%%%%%%%%%%%%%%%%%%%%%%%%%%%%%%%%%%%%%%%%%%%%%%%%%%%%%%%%%%%%%%%%%%
%%%%%  Introduction  %%%%%%%%%%%%%%%%%%%%%%%%%%%%%%%%%%%%%%%%%%%%%%%%%%%%%%%%%%%%%%%%
%%%%%%%%%%%%%%%%%%%%%%%%%%%%%%%%%%%%%%%%%%%%%%%%%%%%%%%%%%%%%%%%%%%%%%%%%%%%%%%%%%%%%

\section{Introduction}
\label{introduction}

\begin{figure}
 \begin{center}
  \includegraphics[angle=0,width=0.9\linewidth]{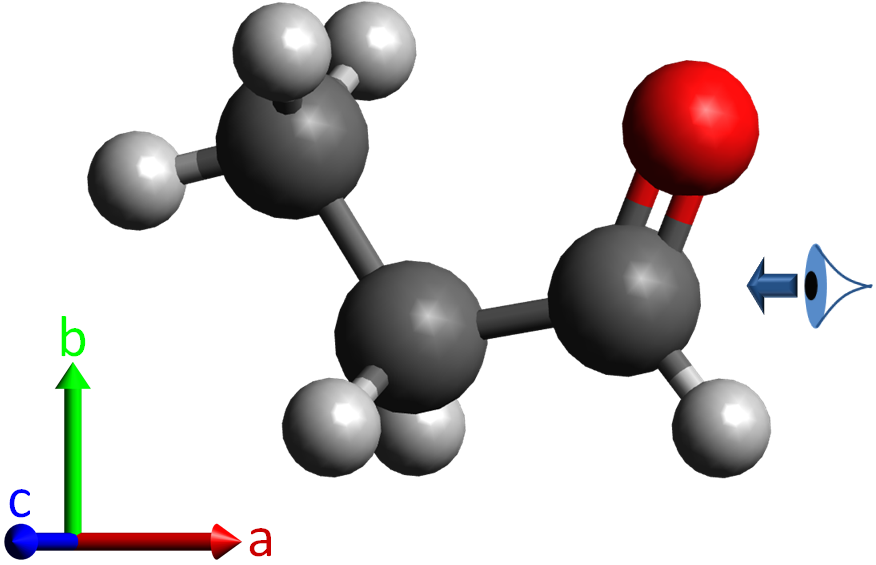}
 \end{center}
  \caption{Sketch of \textit{syn} propanal, CH$_3$CH$_2$CHO. White spheres designate hydrogen, black ones carbon, 
  and the red one oxygen atoms. All heavy atoms, three carbon and one oxygen, are in the \textit{ab}-plane.}
  \label{sketch_syn_propanal}
 \end{figure}

\begin{figure}
 \begin{center}
  \includegraphics[angle=0,width=0.9\linewidth]{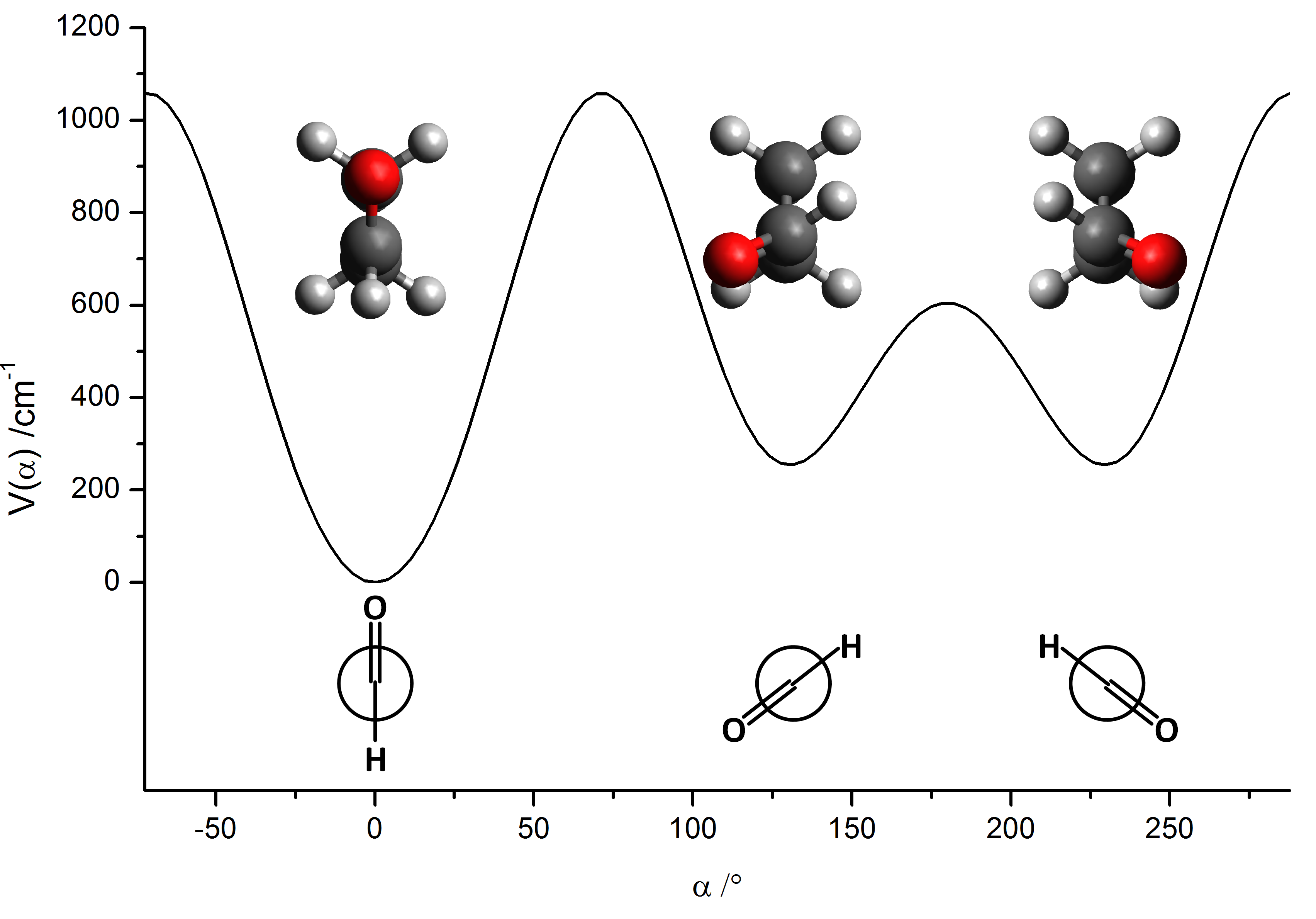}
 \end{center}
  \caption{Calculated potential energy surface of propanal with respect to the rotation 
  of the aldehyde group \cite{propanal_FIR}. There are two stable conformers: \textit{Syn}- and 
  the twofold degenerate \textit{gauche}-conformer. White spheres in propanal sketches designate hydrogen, black ones carbon, 
  and red ones oxygen atoms. The molecular axes \textit{c} and \textit{b} are aligned to the \textit{x}- and \textit{y}-axis, respectively. The point of view is shown for \textit{syn} with the eye in Fig.~\ref{sketch_syn_propanal}. Since \textit{syn} and \textit{gauche} conformers differ mainly by the rotation of the aldehyde group, this alignment was chosen for clarification. Their simplified Newman projections, where the three carbon atoms are shown as one cycle and hydrogen atoms are omitted, except the one from the aldehyde group, are highlighting this rotation. A more illustrative alignment of \textit{syn} is shown in the supplementary material.}
  \label{potential}
 \end{figure}

In 1964, Butcher and Wilson recorded the microwave spectrum of propanal, also known as propionaldehyde, and 
deuterated isotopic species up to 38$\,$GHz. The existence of two stable conformers, \textit{syn} 
(also called \textit{cis}) and the doubly-degenerate \textit{gauche} conformer (from here on simply \textit{syn} and 
\textit{gauche}) was established \cite{propanal_lab_1964}. A sketch of \textit{syn} is shown in Fig.~\ref{sketch_syn_propanal}. The two conformers differ mainly by the rotation of the aldehyde group with respect to 
the carbon atom plane of the molecule. Furtheron transitions in low-lying excited vibrational states were also identified  
\cite{propanal_lab_1964}. The potential energy surface of the rotation of the aldehyde group was studied in detail later 
\cite{propanal_lab_1974,propanal_lab_1988_2}. It was found that \textit{gauche} is higher in energy by 
$\sim$420$\,$cm$^{-1}$ and the aldehyde group is rotated by $\pm$128.2$^\circ$ compared to the \textit{syn} 
orientation \cite{propanal_lab_1988_2}. The potential energy surface of propanal with respect to the rotation 
of the aldehyde group and sketches of the conformers are shown in Fig.~\ref{potential}. 
The torsional potentials were determined more accurately by Far-IR spectroscopy \cite{propanal_FIR} and 
higher-lying vibrational modes were studied by Mid-IR spectroscopy \cite{propanal_IR}. 
In particular, the first aldehyde torsion $\varv_{24} = 1$ is 135.1$\,$cm$^{-1}$ and the first 
excited methyl torsion $\varv_{23} = 1$ is 219.9$\,$cm$^{-1}$ above the ground state of \textit{syn} \cite{propanal_FIR}; 
note that the methyl torsional assignments have been exchanged between \textit{syn} and \textit{gauche} 
from Ref.~\cite{propanal_FIR} to Ref.~\cite{propanal_IR}. In the former work the energy difference between the conformers was
redetermined as $\sim$421$\,$cm$^{-1}$, whereas in the latter work it was determined to be $\sim$370$\,$cm$^{-1}$. 

%%%%%%%%%%%%%%%%%%%%%%%%%%%%%%%%%%%%%%%%%%%%%%%%%%%%%%%%%%%%%%%%%%%%%%%%%%%%%%%%%%%%%
%%%%%  Figure 1a  %%%%%%%%%%%%%%%%%%%%%%%%%%%%%%%%%%%%%%%%%%%%%%%%%%%%%%%%%%%%%%%%%%%%
%%%%%%%%%%%%%%%%%%%%%%%%%%%%%%%%%%%%%%%%%%%%%%%%%%%%%%%%%%%%%%%%%%%%%%%%%%%%%%%%%%%%% 

Whereas the rotational spectrum of \textit{syn} displays only small splittings caused by an internal rotation of the methyl group, 
the barrier between the two equivalent \textit{gauche} conformations is sufficiently low (see Fig.~\ref{potential}) 
such that tunneling between them occurs, causing somewhat larger splittings \cite{propanal_lab_1964}.

%%%%%%%%%%%%%%%%%%%%%%%%%%%%%%%%%%%%%%%%%%%%%%%%%%%%%%%%%%%%%%%%%%%%%%%%%%%%%%%%%%%%%
%%%%%  Figure 1b  %%%%%%%%%%%%%%%%%%%%%%%%%%%%%%%%%%%%%%%%%%%%%%%%%%%%%%%%%%%%%%%%%%%%
%%%%%%%%%%%%%%%%%%%%%%%%%%%%%%%%%%%%%%%%%%%%%%%%%%%%%%%%%%%%%%%%%%%%%%%%%%%%%%%%%%%%% 

Two groups measured further transitions of \textit{syn} independently of each other in the frequency region 
from 8$-$38$\,$GHz using a Stark-modulated and a Fourier transform microwave spectrometer, respectively, but published it jointly \cite{propanal_lab_1982}. 
They used a $S$-type centrifugal distortion analysis to guarantee a correct assignment of transitions involving states 
of large angular momentum \textit{J}. Additionally, the height of the methyl barrier hindering internal rotation in \textit{syn} 
was determined to $\sim$793.7$\,$cm$^{-1}$. 
For \textit{gauche} it is $\sim$886$\,$cm$^{-1}$ and was determined by measuring \textit{c}-type transitions \cite{propanal_lab_1987}. 
This was crucial since \textit{b}- and \textit{c}-type transitions were predicted to split by internal rotation and only \textit{a}-types were assigned before \cite{propanal_lab_1964,propanal_lab_1974}. 
Assignment of rotational transitions for \textit{syn} were extended nearly up to 300$\,$GHz 
with a spectrometer employing super-heterodyne detection \cite{propanal_lab_1987_2}. 
Structural information on either conformer were derived from twelve isotopologues of \textit{syn} and six of 
\textit{gauche} \cite{propanal_lab_1988}. 

In the homologous series of alkanales, the three smallest members, methanal, ethanal, and propanal were already 
detected in space \cite{methanal_det_1969,ethanal_det_1973,propanal_det_2004}. The first detection of propanal was 
towards Sgr B2(N) \cite{propanal_det_2004}. Later, it was detected also in other galactic center molecular clouds 
\cite{propanal_det_2008} and very recently in the Protostellar Interferometric Line Survey (PILS) \cite{Pils} of the low-mass protostellar binary IRAS 16293$-$2422 \cite{propanal_det_2017} using the Atacama Large Millimeter/submillimeter Array (ALMA). 
The first two detections indicated that the molecule is located in colder environments, whereas the most recent one 
is in a comparably warm one at around 125$\,$K. Its detection is in the warm gas close to the protostar, where molecules desorb from the grain ice mantle and the column density is high. The lines toward one component in IRAS 16293$-$2422 are very narrow, approximately 1$\,$kms$^{-1}$ wide, and the PILS data indicated small discrepancies between some observed and 
predicted rotational lines of propanal \cite{propanal_det_2017}.
The discrepancies may arise in part from the complexity 
of the astronomical spectrum itself, especially from presently unaccounted absorption lines affecting the emission lines. 
Furthermore, some of the lines in the observed frequency region (329$-$363$\,$GHz) had non-negligible uncertainties. 
Finally, the identification of propanal in the survey was based on predictions which had ignored the small, 
often unresolved internal rotation splitting. Since higher accuracies are desired, especially in higher frequency regions, 
and propanal was also found in a warmer environment, the spectrum of propanal was measured from 75 up to 500$\,$GHz 
at room temperature with the Cologne (Sub-)Millimeter wave spectrometer and analyzed with respect to the splittings 
of the lines and its energetically lowest lying vibrational states. 
Another incentive for us to revisit the rotational spectrum of propanal was the presence of the two large amplitude 
vibrations and their potential interaction. In the current work, we present results on the ground vibrational states 
of \textit{syn} and \textit{gauche} and the first excited aldehyde and methyl torsional states of \textit{syn}.

%%%%%%%%%%%%%%%%%%%%%%%%%%%%%%%%%%%%%%%%%%%%%%%%%%%%%%%%%%%%%%%%%%%%%%%%%%%%%%%%%%%%%
%%%%%  Experimental details and observed spectrum  %%%%%%%%%%%%%%%%%%%%%%%%%%%%%%%%%%
%%%%%%%%%%%%%%%%%%%%%%%%%%%%%%%%%%%%%%%%%%%%%%%%%%%%%%%%%%%%%%%%%%%%%%%%%%%%%%%%%%%%%

\section{Experimental details}
\label{exptl_details}

The spectrum of propanal was measured continuously in the frequency regions of 75$-$129$\,$GHz and 169.2$-$500.4$\,$GHz with the Cologne (Sub-)Millimeter wave Spectrometer. 
The experimental setup is described in detail elsewhere \cite{cologne_expsetup}. In this work, a clean pyrex cell was used. A rooftop mirror was installed at one end of the 5$\,$m long cell with a diameter of 10$\,$cm to double the absorption path length.
For covering the 75$-$129$\,$GHz range, in-house developed electronics for operating an amplifier tripler chain ($\sim$10$\,$mW) in full saturation mode and a low noise room temperature Schottky detector were used to improve the signal to noise ratio (SNR). 
The higher frequency range was operated with commercial multiplier chains \cite{VDI}.  
Typical background pressures (vacuum) reached 0.1$\,$\textmu bar and the cell was filled with a commercially available sample of propanal (97$\,$\%) from Sigma-Aldrich to a total pressure of about 20$\,$\textmu bar (2$\,$Pa). 
Frequency modulation (FM; $\sim$50\,kHz, amplitude usually between 180 and 300$\,$kHz) was employed with 2\textit{f} demodulation which causes isolated absorption signals to appear close to a second derivative of a Gaussian. The step size varied between 81 and 108$\,$kHz, and the integration time was usually 50$\,$ms. Approximately two gigahertz could be measured per hour.
We assigned uncertainties of 20~kHz to strong and very symmetric lines, 40 or 80~kHz to weaker or less symmetric lines.

%%%%%%%%%%%%%%%%%%%%%%%%%%%%%%%%%%%%%%%%%%%%%%%%%%%%%%%%%%%%%%%%%%%%%%%%%%%%%%%%%%%%%
%%%%%  Spectroscopic analysis  %%%%%%%%%%%%%%%%%%%%%%%%%%%%%%%%%%%%%%%%%%%%%%%%%%%%%%
%%%%%%%%%%%%%%%%%%%%%%%%%%%%%%%%%%%%%%%%%%%%%%%%%%%%%%%%%%%%%%%%%%%%%%%%%%%%%%%%%%%%%

\section{Spectroscopic analysis}
\label{analysis}

\begin{figure}[t]
 \begin{center}
  \includegraphics[angle=0,width=0.9\linewidth]{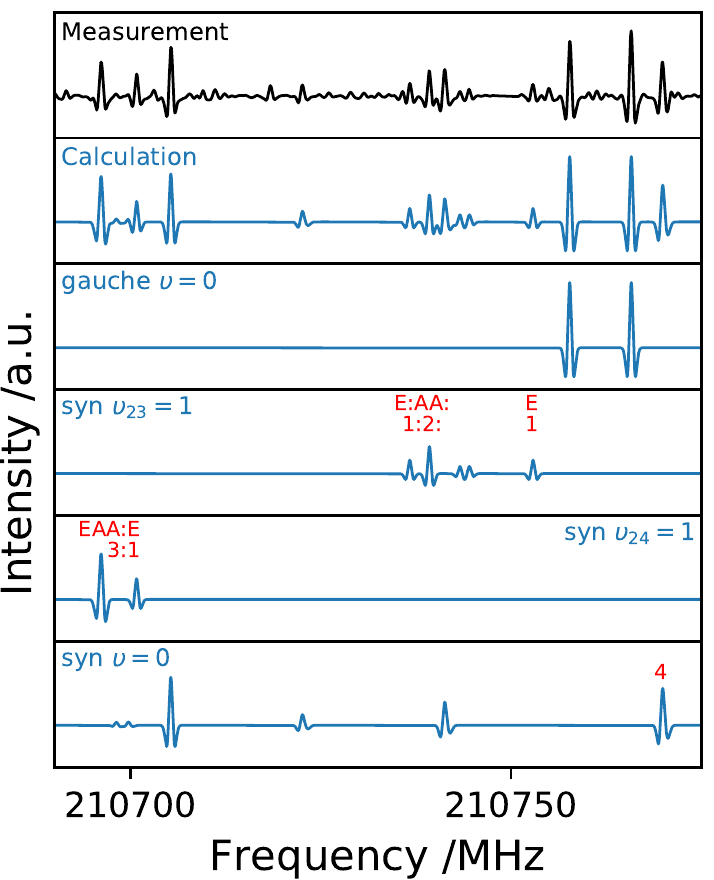}
 \end{center}
  \caption{In the top panel is the measured spectrum of propanal around 210.7$\,$GHz in black in order to show various contributions of the states studied in this work. Beneath is in blue the calculated spectrum including, also shown separately in blue from the bottom panel upwards, synthetic spectra of the ground state ($\varv = 0$), first excited aldehyde ($\varv _{24} = 1$) as well as first excited methyl torsion ($\varv _{23} = 1$) of \textit{syn} and the ground state of \textit{gauche}. Resulting A and E lines due to internal rotation in \textit{syn} can be blended or be separated. For \textit{R}-type \textit{a}-type transitions additional asymmetry splitting with $K_a+K_c=J$ and $K_a+K_c=J+1$ could be observed. These 4 transitions are frequently blended for higher $J$ values. Exemplary (in red) different intensity ratios are shown. In the ground state all 4 lines are blended for $J'=20$ and $K_a=17$, in $\varv _{24} = 1$ a 3:1 ratio is observed for $J'=20$ and $K_a=10$, and a 1:2:1 ratio for $\varv _{23} = 1$ for $J'=20$ and $K_a=11$. Asymmetric lineshapes appear if transitions are not completely merged, yet unresolved. Unassigned lines belong to further vibrationally excited states, isotopologues and possible impurities in the cell.}
  \label{overview_spectrum_propanal}
 \end{figure}

The rotational spectrum of propanal is very rich even at the lowest frequencies of our study. 
In Figure \ref{overview_spectrum_propanal}, observed lines of ground vibrational states of \textit{syn} and \textit{gauche} 
and the first excited aldehyde and methyl torsional states of \textit{syn} are shown. Even though spectra are quite dense, unambiguous assignments can be made. In general, intensity ratios and trends in deviations of observed and calculated lines for different series were used to unambiguously assign transitions to the observed lines.
In fact, in the higher frequency region, this high density of lines is restraining our assignments for lines with large uncertainties more often than limitations in the signal-to-noise 
ratio. Figures~\ref{overview_spectrum_propanal}~and~\ref{gauche_series} already give a hint of the line confusion, however they were choosen to clarify different aspects of the spectra, so there are much more crowded frequency regions. Thus far, about 15\,\% of the observed lines could be assigned 
to the ground states of both conformers and to the two lowest vibrationally excited states of \textit{syn}.

%%%%%%%%%%%%%%%%%%%%%%%%%%%%%%%%%%%%%%%%%%%%%%%%%%%%%%%%%%%%%%%%%%%%%%%%%%%%%%%%%%%%%
%%%%%  Figure 2  %%%%%%%%%%%%%%%%%%%%%%%%%%%%%%%%%%%%%%%%%%%%%%%%%%%%%%%%%%%%%%%%%%%%
%%%%%%%%%%%%%%%%%%%%%%%%%%%%%%%%%%%%%%%%%%%%%%%%%%%%%%%%%%%%%%%%%%%%%%%%%%%%%%%%%%%%%

%%%%%%%%%%%%%%%%%%%%%%%%%%%%%%%%%%%%%%%%%%%%%%%%%%%%%%%%%%%%%%%%%%%%%%%%%%%%%%%%%%%%%
\subsection{\textit{syn}-propanal}
\label{syn}
%%%%%%%%%%%%%%%%%%%%%%%%%%%%%%%%%%%%%%%%%%%%%%%%%%%%%%%%%%%%%%%%%%%%%%%%%%%%%%%%%%%%%

The lower energy \textit{syn} conformer is an asymmetric top rotor with $\kappa = -0.7855$ 
in the ground vibrational state (prolate case: $\kappa = -1$). 
The dipole moment components of \textit{syn} were determined from Stark-effect measurements as 
1.71$\,$D and 1.85$\,$D along the \textit{a}- and \textit{b}-axis, respectively \cite{propanal_lab_1964}. 
The $c$-component is zero owing to the $C_S$-symmetry of the conformer. Rotational lines may be split 
into $A$ and $E$ components with equal intensities due to the threefold degenerate potential 
of the methyl group for internal rotation. A full description of molecular symmetry, 
with symmetry tables for the rigid rotor assumption and allowing for internal rotation, as well as 
resulting energy levels and selection rules can be found in the supplementary material.

In general, for all studied \textit{syn} states, lines in a frequency window of $\sim$30$\,$GHz were assigned and then a new fit was performed, 
if necessary with additional parameters. In general, intensity ratios and trends in deviations
 of observed and calculated lines for different series were used to unambiguously assign transitions 
 to the observed lines. With improved parameters, a prediction for the next frequency window 
was made and so on. Indeed, for both vibrationally excited states only limited microwave data were available. Thus assignments in the millimeter wave range 
were more challenging. Due to no overlapping frequency regions from literature and our spectra, centrifugal distortion constants were kept fixed to the ground state values as a starting point. In a first step, the 
rotational parameters $A$, $B$, $C$, and the energy tunneling parameter $\epsilon_{10}$ were fit 
to the literature data. Subsequently, strong and well separated lines could be assigned for the first few gigahertz. 
Improved centrifugal distortion constants led to improved predictions especially for lines slightly higher 
in $J$ or in $K_a$. 

Groner's ERHAM program \cite{erham} was used for predicting and fitting of the rotational spectrum, 
including splittings due to internal rotation (for the ground vibrational state an extended version was used, since the original version is limited to 8191 transitions). To calculate an initial unitless internal rotation parameter 
$\rho$ and the angle $\Theta_{RAM}$ for applying the Rho-Axis Method (RAM), the formulas 
$\rho_g=\lambda_gI_\alpha/I_g$ and $\Theta_{RAM}=arctan(\rho_b/\rho_a)$ were used ($g=a,b,c$) \cite{Kleiner_rho}. 
The direction cosines of the internal rotation axis $i$ of the top in the principal axis system $\lambda_g$, 
moments of inertia $I_g$, and moment of inertia of the top $I_\alpha$ were taken or calculated with 
rotational parameters and the angle $\sphericalangle(a,i)$ from the literature \cite{propanal_lab_1982}, 
resulting in $\rho=0.06$ and $\Theta_{RAM}=28^\circ$.

The ground vibrational state shows only small, whereas excited vibrational states show larger splittings.
Hence, the ratio of observed lines to assigned transitions is higher for vibrationally excited states 
(specifically, ground state transitions are more frequently blended) and more tunneling parameters 
are required to reproduce the data within experimental uncertainties. In Table \ref{parameters_syn}, 
rotational and tunneling parameters of the ground state and first excited aldehyde-torsion as well as 
first excited methyl-torsion of \textit{syn} are listed.

%%%%%%%%%%%%%%%%%%%%%%%%%%%%%%%%%%%%%%%%%%%%%%%%%%%%%%%%%%%%%%%%%%%%%%%%%%%%%%%%%%%%%
\subsubsection{the ground state, $\varv = 0$ ($E=0\,$cm$^{-1}$)}
\label{GS}
%%%%%%%%%%%%%%%%%%%%%%%%%%%%%%%%%%%%%%%%%%%%%%%%%%%%%%%%%%%%%%%%%%%%%%%%%%%%%%%%%%%%%

Starting with the ground state of \textit{syn}-propanal at 75$\,$GHz, assignments were more or less straightforward, 
since some transitions were already measured and assigned up to 290$\,$GHz. Great care had to be taken into account 
with some typos and bad lines from the literature, conspicuously large deviations from experimental to calculated frequencies. For example, the $A$ and $E$ components of transition $6_{4,~2}-5_{3,~3}$ were listed at 145035.5$\,$GHz, whereas we predict this line at 143035.5$\,$GHz, shifted by 2$\,$GHz. Since the bad lines were unknown at this point and due to a large overlap from literature data and ours, transitions from Ref. \cite{propanal_lab_1987_2} were completely excluded from the fit after first assignments were done in the millimeter wave range. In the end, these 112 lines (with corrected typos) were included in an additional fit and only four show larger deviations, see supplementary material. One of them is the line mentioned above, for the other three lines the splitting due to internal rotation was neglected.
Due to comparably strong dipole moment components along
the \textit{a}- and \textit{b}-axis, mostly \textit{a}- and \textit{b}-type \textit{R}- and \textit{Q}-type transitions were assigned ($a-R$: 2963, $a-Q$:~542, $b-R$:~1567, $b-Q$:~2832, $b-P$:~4). ERHAM also labels some transitions as false (arising from mislabeling) or "forbidden" (mixed wavefunctions, i.e., when the tunneling splitting is approximately equal to the torsional splitting) \textit{c}-type \textit{R-} (186), \textit{Q-} (163), and \textit{P}-type (23) transitions plus 271 so called x-type \textit{P}-type transitions, see Ref. \cite{dimethyl_ether}. In general, energy levels with higher angular momentum $J$ show smaller splittings for identical $K$'s. The $b$-type transitions with low $J$ and high $K_a$ are frequently split; splitting in $a$-type transitions was resolved for $R$-type transitions with large asymmetry splitting ($2K_a \approx J$) and lower values of $J$. 
The root mean square (RMS) error of the fit 
was quite constant up to 300$\,$GHz. The deterioration of the fit quality with higher frequency assignments, up to $\sim$20\,\%, 
was ameliorated with the inclusion of $L_{KKJ}$. 

%%%%%%%%%%%%%%%%%%%%%%%%%%%%%%%%%%%%%%%%%%%%%%%%%%%%%%%%%%%%%%%%%%%%%%%%%%%%%%%%%%%%%
\subsubsection{the 1$^{st}$ excited aldehyde torsion, $\varv_{24} = 1$ ($E=135.1\,$cm$^{-1}$)}
\label{aldehyde_torsion}
%%%%%%%%%%%%%%%%%%%%%%%%%%%%%%%%%%%%%%%%%%%%%%%%%%%%%%%%%%%%%%%%%%%%%%%%%%%%%%%%%%%%%

In general, assignments and fitting were nearly straightforward for the first excited aldehyde torsion, similar to the ground state. 
Noticeable is that series 
of \textit{b}-type transitions, with higher $K_a$ quantum numbers than were already implemented in the fit, needed 
to be weighted carefully step by step into the dataset. Sometimes, this was also observed in the ground state, 
albeit to a lesser extent. With a sufficient amount of new lines these transitions were fit satisfactorily, 
and assignments in the submillimeter wave region could be made. Eventually, all assigned lines from the first excited 
aldehyde torsion could be fit within experimental uncertainty.

%%%%%%%%%%%%%%%%%%%%%%%%%%%%%%%%%%%%%%%%%%%%%%%%%%%%%%%%%%%%%%%%%%%%%%%%%%%%%%%%%%%%%
\subsubsection{the 1$^{st}$ excited methyl torsion, $\varv_{23} = 1$ ($E=219.9\,$cm$^{-1}$)}
\label{methyl_torsion}
%%%%%%%%%%%%%%%%%%%%%%%%%%%%%%%%%%%%%%%%%%%%%%%%%%%%%%%%%%%%%%%%%%%%%%%%%%%%%%%%%%%%%

Assignment and fitting of transitions within the first excited methyl torsion of \textit{syn} turned out to be much more complicated. 
First, observed lines are weaker, so lines are more frequently blended and, due to many weak lines in the spectra, 
special care had to be taken into account to avoid misassignments. Furthermore, larger splittings due to internal rotation 
hampered assignments by finding patterns. We have not been able to fit all assigned lines within experimental uncertainties. 
Nevertheless, out of the 3688 assigned transitions 2760 were reproduced satisfyingly. The observed deviations may originate 
from incorrect assignments, however, the discontinuities in our dataset are evidence of perturbed rotational energy levels. 
For example, an oblate pairing of transitions is observed for high $J$ values with $K_a=3,~4$, and lines can be assigned 
up to $J_{max}=51$. In the final fit lines with $J=39-46$ are excluded and show deviations up to 5$\,$MHz.

The present model treats the $\varv_{23} = 1$ ($A''$ symmetry) state as isolated. However, its energy of 219.9~cm$^{-1}$ 
\cite{propanal_FIR,propanal_IR} is not only comparable to $\varv_{24} = 2$ at 269.3~cm$^{-1}$, the second excited state 
of the aldehyde torsion ($A'$), but also to the $A'$ CCC-bending fundamental $\varv_{15} = 1$ at 264.0~cm$^{-1}$ 
\cite{propanal_FIR,propanal_IR}. Thus, Coriolis interaction may occur between $\varv_{23} = 1$ and $\varv_{15} = 1$. 
In addition, rotational or Coriolis-type interaction may occur between $\varv_{23} = 1$ and $\varv_{24} = 2$, 
which is most likely less prominent because it is of higher order than the Coriolis interaction between two 
fundamentals, besides, $\varv_{24} = 2$ is farther away from $\varv_{23} = 1$ than is $\varv_{15} = 1$. 
The analysis likely becomes even more complex because of a possible Fermi resonance between $\varv_{15} = 1$ and 
$\varv_{24} = 2$.

As usual, the value of $J_{max}$ used in the fit is almost always decreasing for transitions involving energy levels with increasing $K_a$ quantum numbers. 
For $K_a=0$, $J_{max}$ is 54, whereas for $K_a=16$ it is 19. Excluded lines with higher quantum numbers are still 
unweighted in the fit file, but need to be treated with caution, considering that a proper treatment of interactions 
could lead to new assignments, since frequencies can be shifted up to several megahertz, and the spectra are crowded 
with lines. In this case, a new fit was performed by starting from scratch and increasing $J$ and $K_a$ step by step, with no improvement. 
A full treatment of interactions could enable to include all assigned lines in the future.

%%%%%%%%%%%%%%%%%%%%%%%%%%%%%%%%%%%%%%%%%%%%%%%%%%%%%%%%%%%%%%%%%%%%%%%%%%%%%%%%%%%%%
%%%%%  Table 1   %%%%%%%%%%%%%%%%%%%%%%%%%%%%%%%%%%%%%%%%%%%%%%%%%%%%%%%%%%%%%%%%%%%%
%%%%%%%%%%%%%%%%%%%%%%%%%%%%%%%%%%%%%%%%%%%%%%%%%%%%%%%%%%%%%%%%%%%%%%%%%%%%%%%%%%%%%

\begin{table*}
\begin{center}

  \caption{Spectroscopic parameters$^a$ (MHz) of the ground state ($\varv = 0$), first excited aldehyde 
  ($\varv _{24} = 1$) as well as first excited methyl torsion ($\varv _{23} = 1$) of \textit{syn}-propanal.}
  \label{parameters_syn}
{\footnotesize
  \begin{tabular}{llr@{}lr@{}lr@{}l}
  \hline
   & Parameter & \multicolumn{2}{c}{$\varv = 0$} & \multicolumn{2}{c}{$\varv _{24} = 1$} 
   & \multicolumn{2}{c}{$\varv _{23} = 1$} \\
  \hline
&$A$                           & 16669&.626420~(154)  & 16712&.057839~(167) & 16642&.87187~(61)  \\
&$B$                           &  5893&.503711~(49)  & 5857&.196275~(49) & 5871&.180340~(171)  \\
&$C$                           &  4598&.982111~(48)  & 4595&.250184~(50) & 4592&.798935~(116)  \\
&$\it\Delta_K \times 10^3$             & 50&.06225~(63)  & 49&.70451~(84) & 43&.9445~(37)  \\
&$\it\Delta_{JK} \times 10^3$                      & $-$19&.370540~(153)  & $-$18&.097558~(232) & $-$17&.5153~(34)  \\
&$\it\Delta_J \times 10^3$                         & 5&.354247~(37)  & 5&.236678~(38) & 5&.223946~(211)  \\
&$\delta_K \times 10^3$             & 4&.211646~(164)  & 3&.22843~(30) & 1&.6032~(32)  \\
&$\delta_J \times 10^3$             & 1&.5662438~(66)  & 1&.5132959~(108) & 1&.527602~(94)  \\
&$\it\Phi_K \times 10^9$             & 986&.40~(80)  & 1070&.36~(179) & $-$221&.1~(104)  \\
&$\it\Phi_{KJ} \times 10^9$          & $-$606&.15~(37)  & $-$577&.21~(64) & 143&.2~(109)  \\
&$\it\Phi_{JK} \times 10^9$          & 77&.271~(65)  & 62&.316~(141) & $-$128&.2~(92)  \\
&$\it\Phi_J \times 10^9$             & 2&.1239~(88)  & 2&.0061~(92) & 0&.827~(124)  \\
&$\phi_K \times 10^9$             & 292&.50~(58)  & 0&.03746~(138) & $-$1&.471~(46)  \\
&$\phi_{JK} \times 10^9$             & 27&.236~(49)  & 25&.349~(112) & 138&.8~(57)  \\
&$\phi_J \times 10^9$             & 1&.49907~(138)  & 1&.28486~(267) & 0&.555~(62)  \\
&$L_{K} \times 10^{12}$           & &$-$  & $-$79&.43~(120) & &$-$  \\
&$L_{KKJ} \times 10^{12}$         & 14&.082~(239)  & 21&.93~(41) & &$-$  \\
&$L_{JK} \times 10^{12}$          & &$-$  & $-$4&.736~(160) & &$-$  \\
&$L_{JJK} \times 10^{12}$         & &$-$  & &$-$ & $-$71&.7~(56)  \\
&$l_{KJ} \times 10^{12}$         & &$-$  & &$-$ & $-$809&~(32)  \\
&$l_{JK} \times 10^{12}$         & &$-$  & &$-$ & $-$16&.49~(272)  \\
                             & &              &       &              &       &              \\
&$\epsilon_{10}$                                 & $-$3&.0942~(57)   & 38&.7984~(68) & 130&.5541~(250)  \\
&$\epsilon_{20}\times 10^3$                                 & &$-$   & &$-$ & 68&.4~(71)  \\
&$\left[A-(B+C)/2\right]_{10}\times 10^3$         &  &$-$     & $-$2&.477~(33) & $-$7&.887~(224) \\
&$\left[(B+C)/2\right]_{10}\times 10^3$         &  &$-$     & $-$0&.7392~(62) & $-$3&.761~(38) \\
&$\left[(B-C)/4\right]_{10}\times 10^3$          & 0&.09286~(194)      & $-$1&.1251~(80) & $-$3&.8450~(229)  \\
&$\it\Delta_{K_{10}} \times 10^6$             &   &$-$     & &$-$ & $-$10&.80~(65) \\
&$\it\Delta_{JK_{10}} \times 10^6$             &   &$-$     & 1&.1215~(216) & 8&.83~(38) \\
&$\delta_{J_{10}} \times 10^6$             &   &$-$     & 0&.1903~(33) & 0&.4833~(83) \\
&$d_{2_{10}} \times 10^6$             &   &$-$     & 0&.26299~(269) & 0&.5194~(56) \\
&$h_{1_{10}} \times 10^{12}$             &   &$-$     & $-$29&.11~(97) & &$-$ \\
&$h_{2_{10}} \times 10^{12}$             &   &$-$     & $-$16&.082~(245) & &$-$ \\

&$G_{z_{10}}\times 10^3$                           &  &$-$     & 14&.82~(31) & 67&.0~(33) \\
                              & &              &       &              &       &              \\
&$\rho^b \times 10^3$                           & 65&.681~(116)      & 63&.8032~(90) & 61&.9026~(102)  \\ 
&$\Theta_{RAM} [^\circ]$                        & 27&.871~(60)     & 28&.5299~(48) & 28&.4352~(82)  \\

%  \hline
%Literature& no. of lines              &    &              &    a&              &   a&          \\
%& no. of transitions              &    &              &    a&              &   a&          \\
%& rms error$^b$                 &      1&.0          &      1&.0          &      1&.0          \\
%    \hline
%  \hline
%symmetric lines & no. of lines              &    a&              &    a&              &   a&          \\
%$\Delta\nu=$20$\,$kHz & no. of transitions              &    a&              &    a&              &   a&          \\
%& rms error$^b$                 &      1&.0          &      1&.0          &      1&.0          \\
%  \hline
%asymmetric lines & no. of lines              &    a&              &    a&              &   a&          \\
%$\Delta\nu=$40$\,$kHz & no. of transitions              &    a&              &    a&              &   a&          \\
%& rms error$^b$                 &      1&.0          &      1&.0          &      1&.0          \\
%  \hline
%overlapped lines& no. of lines              &    a&              &    a&              &   a&          \\
%$\Delta\nu=$80$\,$kHz & no. of transitions              &    a&              &    a&              &   a&          \\
%& rms error$^b$                 &      1&.0          &      1&.0          &      1&.0          \\
    \hline
  \hline
Combined fit & no. of lines              &    3901&              &    5084&              &   2085&          \\
& no. of transitions              &    8551&              &    6823&              &   2760&          \\
& rms error$^b$                 &      0&.94          &      0&.93          &      1&.06          \\
    \hline    
  \end{tabular}\\[2pt]
}
\end{center}

$^a$\footnotesize{Watson's $A$ reduction in the $I^r$ representation was used in ERHAM. Numbers in parentheses 
     are one standard deviation in units of the least significant figures. In the first block are rotational 
	 constants, in the second ERHAM tunneling parameters (in Watson's $S$ reduction), and in the last are 
	 general information about the fits.}\\ 
$^b$\footnotesize{Weighted unitless value for the entire fit.}\\
\end{table*}
%%%%%%%%%%%%%%%%%%%%%%%%%%%%%%%%%%%%%%%%%%%%%%%%%%%%%%%%%%%%%%%%%%%%%%%%%%%%%%%%%%%%%
%%%%%%%%%%%%%%%%%%%%%%%%%%%%%%%%%%%%%%%%%%%%%%%%%%%%%%%%%%%%%%%%%%%%%%%%%%%%%%%%%%%%%
%%%%%%%%%%%%%%%%%%%%%%%%%%%%%%%%%%%%%%%%%%%%%%%%%%%%%%%%%%%%%%%%%%%%%%%%%%%%%%%%%%%%%

%%%%%%%%%%%%%%%%%%%%%%%%%%%%%%%%%%%%%%%%%%%%%%%%%%%%%%%%%%%%%%%%%%%%%%%%%%%%%%%%%%%%%
\subsection{\textit{ground state of gauche}-propanal, $\varv = 0$ ($E\approx370\,$cm$^{-1}$)}
\label{gauche}
%%%%%%%%%%%%%%%%%%%%%%%%%%%%%%%%%%%%%%%%%%%%%%%%%%%%%%%%%%%%%%%%%%%%%%%%%%%%%%%%%%%%%

The higher energy \textit{gauche} conformer is an asymmetric top rotor with $\kappa = -0.9849$ 
quite close to the limiting prolate case. It is doubly-degenerate with $C_1$ symmetry. 
The low barrier between the two equivalent minima facilitates tunneling between them, 
lifting the degeneracy and creating a symmetric and an antisymmetric tunneling state 
which are often designated as $0^+$ and $0^-$, respectively. Since a non-vanishing transition 
dipole matrix element $\langle\psi_i\mid\mu_x\mid\psi_j\rangle$ is mandatory to observe 
a rotational transition, the wave functions $\psi_i$ and $\psi_j$ need to have the same 
symmetry for \textit{a}- and \textit{b}-type transitions ($x=a,b$). Hence, these transitions only occur 
within substates $0^+$ and $0^-$. However, the direction of $\mu_c$ is changing for the left and right handed 
version, for that reason \textit{c}-type transitions occur between substates. The dipole moment components 
were determined as  $\mu_a=2.645(5)\,$D, $\mu_b=0.417(6)\,$D, and $\mu_c=1.016(3)\,$D, respectively 
\cite{propanal_lab_1988_2}. The rotational spectrum of \textit{gauche} with its Coriolis-type interactions 
was fit and predicted with Pickett's SPFIT and SPCAT programs \cite{spfit_1991} and using Pickett's 
reduced axis system \cite{RAS-Pickett}. We determined linear combinations of spectroscopic parameters 
$X^+$ and $X^-$ from the two substates, explicitly $X=(X^++X^-)/2$ and $\Delta X=(X^+-X^-)/2$, see, e.g., 
Refs.~\cite{eglyc_1_2003,EtSH_2016}. 

The two tunneling states may interact via Coriolis-type interaction of $a$- or $b$-type symmetry with the leading 
coefficients $F_{bc}$ and $F_{ac}$, respectively. The former connects in particular the upper energy asymmetry 
component of a given $J$ and $K_a$ with the lower energy asymmetry component of the same $J$ and $K_a$. 
The strongest interactions occur for $K_a=1$, $K_a=2$, $K_a=3$, $K_a=4$, and $K_a=5$ at $J=2$, $J=10$, 
$J=22$, $J=36$, and $J=52$, respectively. Transitions involving these states are well described in our model. 
Due to a mixing of energy levels from $0^+$ and $0^-$ substates, also formal \textit{x}-type transitions 
($\Delta K_a$ and $\Delta K_c$ even) between these states are observed. The operator with the $F_{ac}$ 
coefficient may connect levels of the same $J$ with $K_a$ and $K_c$ differing by an odd number. 
We did not have clear evidence for this type of interaction.

A splitting due to internal rotation of the methyl group was not observed in the present investigation. 

Unfortunately, only \textit{a}-type transitions could be assigned with confidence in our observed spectra, 
since intensities are proportional to the square of the dipole moment, and \textit{gauche} is in general 
less populated than \textit{syn} owing to Boltzmann statistics. The \textit{b}-type transitions are by far too weak 
to be observed, but $c$-type transitions may have sufficient intensity to identify them in the spectrum. 
We found evidence for a \textit{Q}-type series with $K'_a-K''_a=5-4$, but the intensities of these lines 
were in some occasions too weak to assign them securely. Two \textit{Q}-type \textit{b}-type 
and one \textit{R}-type \textit{c}-type transition from the literature are unweighted in the fit due to conspicuously 
large deviations.   

%%%%%%%%%%%%%%%%%%%%%%%%%%%%%%%%%%%%%%%%%%%%%%%%%%%%%%%%%%%%%%%%%%%%%%%%%%%%%%%%%%%%%
%%%%%  Figure 3  %%%%%%%%%%%%%%%%%%%%%%%%%%%%%%%%%%%%%%%%%%%%%%%%%%%%%%%%%%%%%%%%%%%%
%%%%%%%%%%%%%%%%%%%%%%%%%%%%%%%%%%%%%%%%%%%%%%%%%%%%%%%%%%%%%%%%%%%%%%%%%%%%%%%%%%%%% 

 \begin{figure}
 \begin{center}
  \includegraphics[angle=0,width=0.9\linewidth]{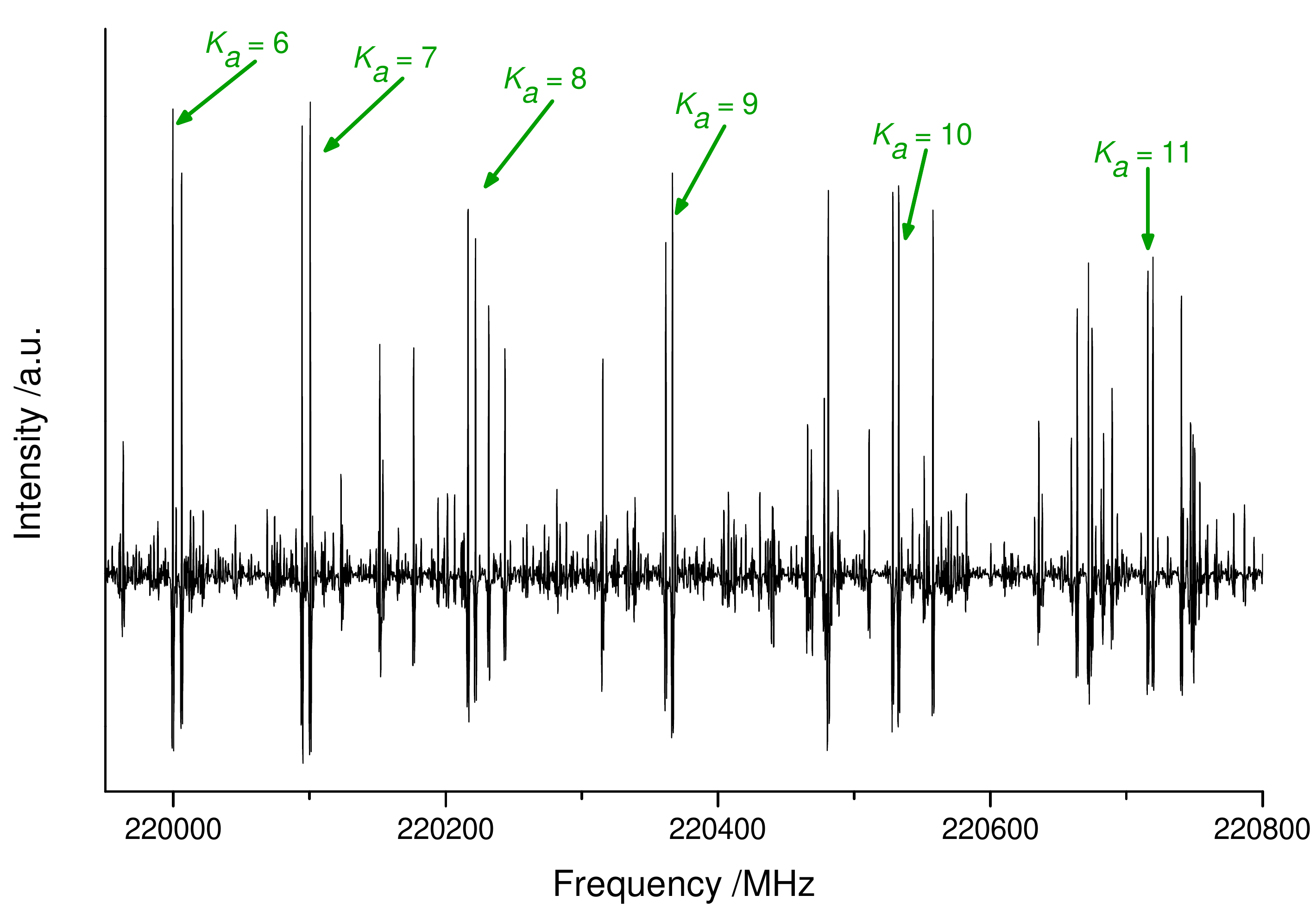}
 \end{center}
  \caption{The measured spectrum of propanal around 220$\,$GHz. Labeled here is a set of doublets 
originating from tunneling rotation interaction from two stable degenerate \textit{gauche}-conformers. The \textit{R}-type \textit{a}-type transitions $26_{K_a,~K_c}-25_{K_a,~K_c-1}$ are marked with their respective $K_a$ quantum numbers. The asymmetry splitting ($J=K_a+K_c$ and $J+1=K_a+K_c$) is unresolved. The intensities are decreasing with increasing $K_a$'s. The strong lines illustrate the possibility of secure assignments of transitions even with higher $K_a$ quantum numbers. Furtheron the density of lines in the measured spectrum is visible, even though this frequency region was choosen for clarification of the doublet pattern, i.e. there are much more crowded regions full of lines with comparable intensities.}
  \label{gauche_series}
 \end{figure}
 
The fitting procedure was very different for \textit{gauche} compared to \textit{syn}. We started to fit all 
transitions with $K_a=0$ up to $J=45$, followed by series with $K_a=1$, then $K_a=2$, and so on. 
Later, transitions involving $J$ up to 60 were added. To assign and especially to fit transitions, 
originating in strongly perturbed energy levels, was quite challenging. For $K_a=2$, transitions from 
unperturbed energy levels with higher \textit{J} quantum numbers were implemented earlier in the fit than 
those from perturbed ones. Otherwise the quality of the fit was too poor. It is possible that the fit converged into 
a different local minimum. In Figure \ref{gauche_series} an easy assignable set of doublets, 
originating by tunneling rotation interaction, of \textit{R}-type \textit{a}-type transitions with $J'-J''=26-25$ is shown.  
Complications arose for transitions with $K_a=12$, which should not be affected by 
Coriolis-type interactions. Here, the dataset could not be fit to the experimental uncertainties for $J>38$, 
even though more transitions can be assigned securely still straightforwardly (517 in total). The strong lines in Fig. \ref{gauche_series} illustrate the possibility of secure assignments to higher $K_a$ quantum numbers, up to $K_a=22$.
The observed deviations could originate from insufficiencies in our spectroscopic model or from interactions with 
other states, for example the first excited aldehyde torsion of \textit{gauche}, and are shown exemplary for $K_a=12$ 
in the supplementary material. The analysis of this excited state 
is beyond the scope of the present work, but is intended for a future study. 
A model with assigned and fitted \textit{a}-type transitions up to 
$J=60$ and $0\leq K_a\leq 13$, except for $J>50$ for $K_a=11$ and $J>38$ for $K_a=12$, is now available. 
Additionally, transitions for lower $J$'s up to 31 with $K_a$ up to 17 are included as far as the RMS error 
was not increasing considerably. Our best choice of rotational and Coriolis interaction parameters for \textit{gauche} 
is listed in Table \ref{parameters_gauche}.

%%%%%%%%%%%%%%%%%%%%%%%%%%%%%%%%%%%%%%%%%%%%%%%%%%%%%%%%%%%%%%%%%%%%%%%%%%%%%%%%%%%%%
%%%%%  Table 2   %%%%%%%%%%%%%%%%%%%%%%%%%%%%%%%%%%%%%%%%%%%%%%%%%%%%%%%%%%%%%%%%%%%%
%%%%%%%%%%%%%%%%%%%%%%%%%%%%%%%%%%%%%%%%%%%%%%%%%%%%%%%%%%%%%%%%%%%%%%%%%%%%%%%%%%%%%

\begin{table}
\begin{center}

  \caption{Spectroscopic parameters$^a$ (MHz) of the two substates for the ground state of \textit{gauche}-propanal.}
  \label{parameters_gauche}
{\footnotesize
  \begin{tabular}{lr@{}lr@{}l}
  \hline
   Parameter & \multicolumn{2}{c}{$X$} & \multicolumn{2}{c}{$\Delta X$} \\
  \hline
$E$      &    & & 237&.7953~(98)  \\ 
$A$                           & 26250&.6640~(148) & 2&.0345~(63)  \\                         
$B$                           & 4314&.92050~(117) & $-$0&.030328~(159)    \\                         
$C$                           & 4148&.01846~(117) & $-$0&.107072~(90)   \\                   
$\it\Delta_K $        & 2&.1000~(41) & $-$0&.008057~(282)   \\
$\it\Delta_{JK}$                      & $-$0&.1794215~(84) & 0&.00046067~(241)   \\
$\it\Delta_J \times 10^3$             & 6&.58936~(136) & $-$0&.008402~(64)   \\
$d_1 \times 10^3$             & 1&.08232~(88) & $-$3&.922~(33)   \\
$d_2 \times 10^6$             & $-$38&.36~(41) & 1&.4260~(195)   \\
$H_{KJ} \times 10^6$          & 9&.613~(70) & 1&.3617~(240)   \\
$H_{JK} \times 10^6$          & $-$3&.3404~(164) & $-$0&.02918~(69)   \\
$H_J \times 10^9$             & 127&.294~(252) & 0&.3408~(245)   \\
$h_1 \times 10^9$             & $-$52&.463~(193) & $-$0&.8260~(54)   \\
$h_2 \times 10^9$             & $-$0&.463~(40) & 0&.4434~(48)   \\
$h_3 \times 10^9$             & $-$0&.1814~(181) & $-$0&.09859~(125)   \\
$L_{KKJ} \times 10^9$         & $-$4&.318~(220) & $-$1&.693~(66)   \\
$L_{JK} \times 10^9$          & $-$2&.263~(73) & 0&.02756~(235)   \\
$L_{JJK} \times 10^{12}$         & 145&.05~(216) & $-$2&.580~(101)   \\
$L_{J} \times 10^{12}$         & $-$3&.248~(33) & 0&.0706~(32)   \\
$l_1 \times 10^{12}$          & 2&.1031~(150) & &$-$   \\
$P_{JK} \times 10^{12}$         & $-$0&.1243~(34) & &$-$   \\
$P_{J} \times 10^{15}$         & 0&.07098~(299) & &$-$   \\
                              & & & &                    \\                             
$F_{bc}$      &    \multicolumn{2}{r@{}}{23.} & \multicolumn{2}{@{}l}{9491~(67)}  \\
$F_{bc{_K}}$      &    \multicolumn{2}{r@{}}{$-$0.} & \multicolumn{2}{@{}l}{1018~(49)}  \\
$F_{bc{_J}} \times 10^3$      &    \multicolumn{2}{r@{}}{$-$3.} & \multicolumn{2}{@{}l}{3778~(62)}  \\ 
$F_{bc{_{KK}}} \times 10^3$      &    \multicolumn{2}{r@{}}{0.} & \multicolumn{2}{@{}l}{23517~(244)}  \\ 
$F_{bc{_{JK}}} \times 10^6$      &    \multicolumn{2}{r@{}}{$-$14.} & \multicolumn{2}{@{}l}{565~(160)}  \\
$F_{bc{_{JJ}}} \times 10^6$     &    \multicolumn{2}{r@{}}{0.} & \multicolumn{2}{@{}l}{14996~(76)}  \\
$F_2{_{bc}} \times 10^3$      &    \multicolumn{2}{r@{}}{$-$0.} & \multicolumn{2}{@{}l}{3618~(39)}  \\
  
\hline
  \hline
no. of lines              &   \multicolumn{2}{r@{}}{1685} & \multicolumn{2}{@{}l}{}             \\
no. of transitions              &    \multicolumn{2}{r@{}}{2378} & \multicolumn{2}{@{}l}{}             \\
rms error$^b$                  &   \multicolumn{2}{r@{}}{0.} & \multicolumn{2}{@{}l}{97}             \\  
    \hline
  \end{tabular}\\[2pt]
}
\end{center}

$^a$\footnotesize{Watson's $S$ reduction was used with SPFIT in the I$^{r}$-representation. Parameters were fit 
   as linear combinations $X=(X^++X^-)/2$ and $\Delta X=(X^+-X^-)/2$. Values of $F_{bc}$ etc., given 
   between two columns, are associated with operators between both tunneling states; $\Delta X$ is not defined 
   for these parameters. Numbers in parentheses are one standard deviation in units of the least significant figures.}\\
$^b$\footnotesize{Weighted unitless value for the entire fit.}
\end{table}

%%%%%%%%%%%%%%%%%%%%%%%%%%%%%%%%%%%%%%%%%%%%%%%%%%%%%%%%%%%%%%%%%%%%%%%%%%%%%%%%%%%%%
%%%%%  Figure 4  %%%%%%%%%%%%%%%%%%%%%%%%%%%%%%%%%%%%%%%%%%%%%%%%%%%%%%%%%%%%%%%%%%%%
%%%%%%%%%%%%%%%%%%%%%%%%%%%%%%%%%%%%%%%%%%%%%%%%%%%%%%%%%%%%%%%%%%%%%%%%%%%%%%%%%%%%%

 \begin{figure}
 \begin{center}
  \includegraphics[angle=0,width=8.5cm]{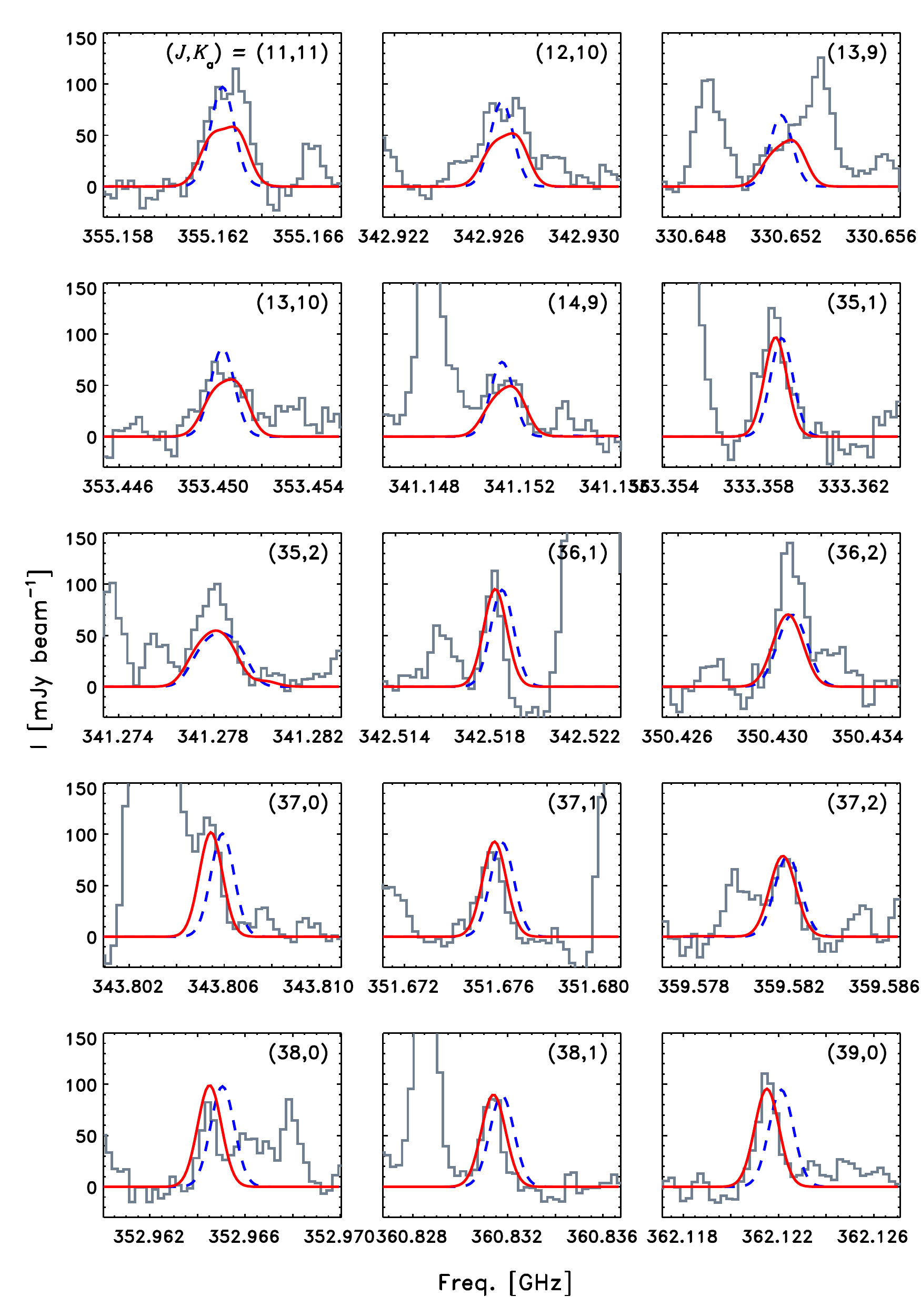}
 \end{center}
  \caption{Sections of the ALMA-PILS radio astronomical spectrum toward a position offset from one of the components in the protostellar binary (see Refs. \cite{Pils} and \cite{propanal_det_2017} for details) displaying selected emission 
           features of propanal. The observed spectrum is shown in solid gray lines, the initial 
		   model is indicated by dashed blue lines, and the new model by solid red lines. 
		   The upper state $J$ and $K_a$ values of the transitions are indicated. The high-$K_a$, 
		   low-$J$ transitions are $b$-type $R$-type transitions with unresolved asymmetry splitting, 
		   the high-$J$, low-$K_a$ transitions are blends of two $a$-type and two $b$-type $R$-type 
		   transitions which all have $K_c = J - K_a$, where $K_a$ is the lower one of two and the one shown.}
  \label{PILS-propanal}
 \end{figure}

%%%%%%%%%%%%%%%%%%%%%%%%%%%%%%%%%%%%%%%%%%%%%%%%%%%%%%%%%%%%%%%%%%%%%%%%%%%%%%%%%%%%%
%%%%%%%%%%%%%%%%%%%%%%%%%%%%%%%%%%%%%%%%%%%%%%%%%%%%%%%%%%%%%%%%%%%%%%%%%%%%%%%%%%%%%
%%%%%%%%%%%%%%%%%%%%%%%%%%%%%%%%%%%%%%%%%%%%%%%%%%%%%%%%%%%%%%%%%%%%%%%%%%%%%%%%%%%%%

%%%%%%%%%%%%%%%%%%%%%%%%%%%%%%%%%%%%%%%%%%%%%%%%%%%%%%%%%%%%%%%%%%%%%%%%%%%%%%%%%%%%%
%%%%%  Discussion  %%%%%%%%%%%%%%%%%%%%%%%%%%%%%%%%%%%%%%%%%%%%%%%%%%%%%%%%%%%%%%%%%%
%%%%%%%%%%%%%%%%%%%%%%%%%%%%%%%%%%%%%%%%%%%%%%%%%%%%%%%%%%%%%%%%%%%%%%%%%%%%%%%%%%%%%

\section{Discussion and conclusion}
\label{Discussion}
The ground and first excited aldehyde torsion of \textit{syn} can be assigned and modeled quite straightforwardly. 
The improvement of centrifugal distortion constants leads to faithful predictions even in higher frequency regions. 
The description of these states with a total of 15374 newly assigned and fitted transitions provides 
satisfactory synthetic spectra for state-of-the-art astronomical observation standards. Only predictions for 
\textit{b}-type transitions with $K_a$ quantum numbers much higher than implemented in the fit should be treated 
with caution, see subsection \ref{aldehyde_torsion}.

Parameters describing the first excited methyl torsion are quite different compared to the other two states. 
Since it was treated as an isolated state, but there is evidence that it is perturbed, the model is adequate 
only for transitions with quantum numbers implemented in the fit. Varying magnitudes of the observed splittings 
due to internal rotation of the methyl group for the various studied states can be explained roughly with 
the classical tunneling scheme: Higher relative energies of vibrationally excited states, consequently, 
smaller energy differences to the potential barrier height of 793.7$\pm$2.5$\,$cm$^{-1}$ lead in general 
to larger splittings. This results in increasing magnitude of the energy tunneling parameter $\epsilon_{10}$, 
from the ground state $\epsilon_{10}=-3.0941~(57)\,$MHz ($E_{rel}=0\,$cm$^{-1}$) to the first aldehyde torsion 
$\epsilon_{10}=38.7984~(68)\,$MHz ($E_{rel}=135.1\,$cm$^{-1}$) up to the first excited methyl torsion 
$\epsilon_{10}=130.5541~(250)\,$MHz ($E_{rel}=219.9\,$cm$^{-1}$).

Modelling of \textit{gauche} is somewhat more difficult. Most of the lines can be assigned confidently, 
but the treatment, even with Coriolis interaction, seems insufficient. The large residuals may be caused by 
perturbations caused by a yet unidentified state. Wrong assignments not yet identified could also be an explanation 
as could be the inadequacy of the effective Hamiltonian, or that the large parameter space causes the fit to 
not converge into the global minimum. Sometimes the fit is converging and the parameter values are reasonable 
with different sets of parameters, so the right choice is not obvious, nor unambiguous. 
Missing \textit{b}- and \textit{c}-type transitions may affect the description of the $K$-level structure, 
and may, in turn, be a reason for our unability to determine $F_{ac}$ with confidence. The magnitude of $F_{ac}$ 
may well be larger than that of $F_{bc}$, as, for example, in ethanethiol \cite{EtSH_2016} or hydroxyacetonitrile 
\cite{HAN_rot_2017}; see also the theoretical treatment of the reduced axis system \cite{RAS-Pickett}. 
The omission of $F_{ac}$ may influence the ability to determine higher order parameters or affect their values. 
Complications by describing molecules with two stable degenerate forms and necessary treatment of Coriolis interactions 
occurred already before \cite{cyanoacetaldehyde_2012,mercaptoacetonitrile_2016}. For conformer I of cyanoacetaldehyde 
only transitions up to $K_a=2$ for $0^+$ and up to $K_a=8$ for $0^-$-state are employed in the fit and moreover transitions 
with $J$ equal to 30, 31, and 32 needed to be excluded \cite{cyanoacetaldehyde_2012}. For \textit{synclinal} mercaptoacetonitrile, 
the standard deviation is four times larger than the expected experimental uncertainty and again special kinds of 
transitions were excluded, regarding large deviations between experimental and calculated frequencies 
\cite{mercaptoacetonitrile_2016}. For \textit{gauche} propanal, a first model is now available into the submillimeter wave region. 

This extensive study of propanal could lead to a better understanding of the behavior of complex molecules with 
two large amplitude motions, here originating from its two functional groups, the aldehyde and methyl groups.

With the newly derived constants of this work the synthetic spectra of \textit{syn} and \textit{gauche} should satisfy 
the requirements in sensitivity and accuracy of astronomical observations nowadays, even for warm environments. 
We have modeled the propanal emission lines with the new predictions and compare them with those previously available. 
As can be seen in Fig.~\ref{PILS-propanal} ($b$-type) transitions with high $K_a$ and low $J$ appear weaker, broader, 
and slightly shifted because of the consideration of the methyl internal rotation in the present predictions. 
Overprediction of line intensities does not occur anymore. Transitions with high $J$ and low $K_a$, on the other hand, 
are shifted by up to half a line-width; whenever the shift is large, the agreement is much better with the new predictions. 

Predictions of the rotational spectra of \textit{syn}- and \textit{gauche}-propanal are 
available in the catalog section\footnote{https://cdms.astro.uni-koeln.de/classic/entries/} of the 
CDMS~\cite{CDMS_1,CDMS_2}. In addition, line, parameter, and fit files, along with other auxiliary files, 
are available in the spectroscopy data 
section\footnote{https://cdms.astro.uni-koeln.de/classic/predictions/daten/CH3CH2CHO/} 
of the CDMS.

%% The Appendices part is started with the command \appendix;
%% appendix sections are then done as normal sections
%% \appendix

%%%%%%%%%%%%%%%%%%%%%%%%%%%%%%%%%%%%%%%%%%%%%%%%%%%%%%%%%%%%%%%%%%%%%%%%%%%%%%%%%%%%%
%%%%%  acknowledgements  %%%%%%%%%%%%%%%%%%%%%%%%%%%%%%%%%%%%%%%%%%%%%%%%%%%%%%%%%%%%
%%%%%%%%%%%%%%%%%%%%%%%%%%%%%%%%%%%%%%%%%%%%%%%%%%%%%%%%%%%%%%%%%%%%%%%%%%%%%%%%%%%%%

\section*{Acknowledgments}

We thank Hanno Schmiedt for group theory discussions and Matthias Ordu for the extended 
version of ERHAM. This work has been supported by the Collaborative Research Centre 956, 
project B3, and the Ger{\"a}tezentrum ``Cologne Center for Terahertz Spectroscopy'', 
both funded by the Deutsche Forschungsgemeinschaft (DFG). OZ acknowledges support 
from the Bonn-Cologne-Graduate School of Physics and Astronomy (BCGS). 
JKJ acknowledges support from the European Research Council (ERC) under the European 
Union's Horizon 2020 research and innovation programme through ERC Consolidator Grant 
``S4F'' (grant agreement No~646908).

\appendix

\section*{Appendix A. Supplementary Material}

Supplementary data associated with this article can be found, in 
the online version, at http://dx.doi.org/10.1016/j.jms.2017.07.008.

%% References
%%
%% Following citation commands can be used in the body text:
%% Usage of \cite is as follows:
%%   \cite{key}         ==>>  [#]
%%   \cite[chap. 2]{key} ==>> [#, chap. 2]
%%

%% References with bibTeX database:

%%%  \bibliographystyle{elsarticle-num}
%%%  \bibliography{<your-bib-database>}

%% Authors are advised to submit their bibtex database files. They are
%% requested to list a bibtex style file in the manuscript if they do
%% not want to use elsarticle-num.bst.

%% References without bibTeX database:

%%%%%%%%%%%%%%%%%%%%%%%%%%%%%%%%%%%%%%%%%%%%%%%%%%%%%%%%%%%%%%%%%%%%%%%%%%%%%%%%%%%%%
%%%%%%%%%%%%%%%%%%%%%%%%%%%%%%%%%%%%%%%%%%%%%%%%%%%%%%%%%%%%%%%%%%%%%%%%%%%%%%%%%%%%%
%%%%%%%%%%%%%%%%%%%%%%%%%%%%%%%%%%%%%%%%%%%%%%%%%%%%%%%%%%%%%%%%%%%%%%%%%%%%%%%%%%%%%

\end{document}